\documentclass[prd,aps]{revtex4}
\usepackage{graphicx}
\newcommand{\bb}{\begin{eqnarray}}
\newcommand{\ee}{\end{eqnarray}}

\begin{document}
\title{ \bf Effect of magnetic fields on bound localized electron 
states}
\author{F.Kh. Chibirova}
\email{chibir@cc.nifhi.ac.ru} \affiliation{Karpov Institute of
Physical Chemistry , 103064, Moscow, Russia}
\author{V.R. Khalilov}
\email{khalilov@thc.phys.msu.su}
\affiliation{Physics Department, Moscow State University,
119899, Moscow, Russia}

\begin{abstract}
An approximate analytic solution for the ground electron state are
found to the Schr\"odinger equation for a combination of a uniform
magnetic field and single attractive $-\lambda\delta({\bf r})$
potential. Effect of the magnetic field on this bound localized electron
state is discussed. We show that above effect leads to appearing
the probability current density in some region centered in the
point ${\bf r}=0$ as well as to enlarging (for a number of physical models)
the localization region of the electron in the plane
perpendicular to the magnetic field. We are expected that these properties
can be of importance for real quantum mechanical systems.
\end{abstract}


\maketitle

\section{Introduction}

When an electron travels in a uniform magnetic field (that is in a
cylindrical external field configuration) where a natural
assumption is that the relevant quantum mechanical system is
invariant along the symmetry ($z$) axis, the system then becomes
essentially two-dimensional in the $xy$ plane \cite{khu}. Many
physical phenomena occurring in quantum systems of electrically
charged fermions, which have the axial symmetry, can be studied
effectively by means of the corresponding Schr\"odinger or Pauli
equation in 2+1 dimensions.

A permanent interest in this topic also is stimulated by the
studies of (2+1)-dimensional models in both superconductivity
\cite{fw} and particle theory (including the quantum Hall effect
\cite{pg} and the degenerate planar semiconductors with low-energy
electron dynamics \cite{ams})[2-4]).

Two-dimensional electrons  in external electromagnetic fields at
low temperature has attracted significant interests since the
discovery of the integral quantum Hall effect (IQHE) \cite{pg}.
The amazing fact that the quantized values of the Hall conductance
are related only to fundamental constants and are independent of
the actual sample and geometry. Another important fact that the
IQHE occurs in impure samples with defect centers when localized
electron states appear. There are defects of such type in thin
films too. It was shown in \cite{chib} that the residual effect
was observed in the M\"ossbauer spectra of defect oxide thin films
after magnetic field treatment.

The purpose of this paper is to investigate the effect of the
magnetic field on planar localized electron states.

\section{Electron in a uniform magnetic field }

At first let us consider an electron of charge $e<0$ in the $xy$
plane in a uniform magnetic field ${\bf B}$, which is specified in
Cartesian coordinates as \bb
 {\bf B}=(0,\,0,\,B)=\nabla\times {\bf A},\ \
 {\bf A}=(-yB,\,0,\,0).
\label{e1} \ee The Pauli equation in field (\ref{e1}) has the form
\bb i\hbar\frac{\partial}{\partial t}\psi({t, \bf r})={\cal
H}\psi({t, \bf r}),\quad {\bf r}=(x, y), \label{eq12} \ee where
the Hamiltonian ${\cal H}$ is \bb {\cal H} =
\frac{1}{2m}\left(-i\hbar\frac{\partial}{\partial
x}+\frac{eB}{c}y\right)^2
-\frac{\hbar^2}{2m}\frac{\partial^2}{\partial y^2} + \mu\sigma_3B.
\label{e12} \ee Here $m>0$ is the effective electron mass in a
crystal lattice, $\mu=|e|\hbar/2m_ec$ is the Bohr magneton, $m_e$
is the mass of a free electron and
$$
\sigma_3=\left(\begin{array}{cc}
1 & 0\\
0 &-1\\
\end{array}\right).
$$
The last term in (\ref{e12})
describes the interaction of the spin magnetic moment of the
electron with the magnetic field.  If the electron is free we
shall put $m=m_e$.

Electron wave function in field (\ref{e1}) has the form
\bb
 \psi_{np}(t, {\bf r})=\frac12 e^{-iE_{ns}t/\hbar}e^{ipx/\hbar}U_n(Y)
\left( \begin{array}{c}
1+s\\
1-s
\end{array}\right)
\label{sol1} \ee where \bb E_{ns}=\hbar \omega
\left(n+\frac12\right)+ s\hbar \omega\frac{m}{2m_e} \label{e2} \ee
is the energy spectrum of two-dimensional electron,
$\omega=|eB|/mc$, $s=\pm 1$ is conserving spin quantum number, $p$
is the momentum of electron in the $x$-direction in field (1).
Note that $p$ is constrained by $|p|\leq eBL/c$ \cite{ll}.

The functions
$$
 U_n(Y) = \frac{1}{(2^n n!\pi^{1/2}a)^{1/2}}
\exp\left(-\frac{(y-y_0)^2}{2a^2}\right)H_n\left(\frac{y-y_0}{a}\right),
$$
are expressed through the Hermite polynomials $H_n(z)$ is the
Hermite polynomial, the integer $n=0, 1, 2, \dots$ indicates the
Landau level number, $a=\sqrt{\hbar c/|eB|}\equiv
\sqrt{\hbar/m\omega}$ is the so-called magnetic length and
$y_0=-cp/eB$. It should be reminded that the classical trajectory
of an electron in the $xy$ plane perpendicular to the magnetic
field is a circle with the rest center. Quantity $y_0$ corresponds
to the classical $y$ - coordinate of the circle center.

All electron states (\ref{sol1}) are not localized in the
$x$-direction.

\section{Electron in an attractive potential in the presence
of a uniform magnetic field}

Now we study a simple solvable model. We consider the motion of an
electron in a single attractive $-\lambda\delta({\bf r})$
potential in the presence of a uniform magnetic field. Here
$\lambda$ is a positive constant, $\delta({\bf r})$ is the Dirac
delta function. In fact, the equation we need solve is the
following Schr\"odinger equation \bb
\frac{1}{2m}\left[\left(-i\hbar\frac{\partial}{\partial
x}+\frac{eB}{c}y\right)^2 -\hbar^2\frac{\partial^2}{\partial y^2}
-\hbar^2\lambda\delta({\bf r})\right]\Psi_E({\bf r})=E\Psi_E({\bf
r}). \label{e21} \ee Indeed, because the electron spin is
conserved, we cannot take into account the last term in
(\ref{e12}). Solutions of Eq.(\ref{e21}) are sought in the form
\bb \Psi_E({\bf r})=\sum\limits_{n=0}^{\infty} \int dp
C_{Enp}\psi_{np}({\bf r}) \equiv \sum\limits_{n,p}
C_{Enp}\psi_{np}({\bf r}), \label{sum1} \ee where $\psi_{np}({\bf
r})$ is the spatial part of the wave functions (\ref{sol1}).

Then for coefficients $C_{Enp}$, it is easily to obtain \bb
C_{Enp}(n+b)=\lambda_0\sum\limits_{l,k}C_{Elk}V_l(0)V_n(0),
\label{syst}\ee where
$$
b=\frac12-\frac{E}{\hbar\omega},\quad \lambda_0=\frac{\lambda}
{4\pi m\omega a},\quad V_l(0)\equiv \sqrt{a}U_l(y=0).
$$ Supposing
 \bb C_{Enp}=C_E\frac{V_n(0)}{n+b},
\label{sup}\ee then inserting (\ref{sup}) in (\ref{syst}) and
taking account of formulas \bb \sum\limits_{n,p}
|C_{Enp}|^2=1,\label{relat}\ee \bb
\sum\limits_{p}V_n(0)V_k(0)\equiv \int dp
V_n(0)V_k(0)=\frac{\hbar}{a}\delta_{n,k},\ee
 we obtain the following equations
\bb 1=\frac{\lambda}{4\pi}\sum\limits_{n=0}^{\infty}\frac{1}{n+b},
\label{ener}\ee \bb (C_E)^{-2}=
\sum\limits_{n=0}^{\infty}\frac{1}{(n+b)^2}. \label{coeff}\ee
Eq.(\ref{ener}) defines implicitly the energy of a bound localized
electron state.

It is seen from Eq. (\ref{ener}) that the electron energy in the
lowest unbound state in the presence of uniform magnetic field is
equal to $\hbar\omega/2$ and the latter keep on to have the same
value in three-dimensional space. Hence, one can conclude that
there always exists a bound electron state in any attractive
potential in the presence of magnetic field in three-dimensional
space. The values of energies certainly differ from each other for
two and three spatial dimensions, nevertheless the motion patterns
in the plane perpendicular to the magnetic field will be like.

It is pleasant feature of two-dimensional model discussed to make
it possible to find both the energy and the approximate wave
function of bound state in clear simple forms. At first it is well
to note that $\lambda$ is a dimensionless constant in
two-dimensional case so, in fact, Eq.(\ref{e21}) does not contain
a term of the energy dimension. Nevertheless, a bound electron
state exists in the attractive $-\lambda\delta({\bf r})$
potential. One can see that the sum on the right of
Eq.(\ref{ener}) diverges. Introducing the cut-off at $n=N$, we
obtain \bb 1=\frac{\lambda}{4\pi}\ln\left(\frac{N}{b}+1\right).
\label{cut} \ee Thus, we have at large $N$ \bb -E+\frac{\hbar
\omega}{2} =\hbar \omega N\exp\left(-\frac{4\pi}{\lambda}\right).
\label{binde} \ee

In order to consider $\lambda$ as the bare coupling constant we
must require that it should depend on $N$ so as to  quantity
$-E+\hbar\omega/2$ would remain finite at $N\to \infty$. Thus, the
cut-off parameter $N$, which tends to infinity, transmutes in
arbitrary binding energy $|E|$. This is a nonrelativistic analog
of the dimensional transmutation phenomenon which occurs in
massless relativistic field theories \cite{colw}. For the model
under discussion in the absence of magnetic field this phenomenon
was first considered in \cite{thor}).

Now omitting simple calculations we give
the final expression for the normalized wave function of bound
electron (see, also \cite{pg1,TeBag}) \bb \Psi_0({\bf
r})=\frac{1}{\sqrt{2\pi}a}\exp\left(\frac{-x^2-y^2+2ixy}{4a^2}\right).
\label{ground} \ee In the plane perpendicular to the magnetic
field electron is localized in the region $x, y\sim a$. This
electron state has a wonderful property. Its probability current
densities are \bb j_x=-\frac{\hbar }{2\pi
ma^4}y\exp\left(\frac{-x^2-y^2}{2a^2}\right)\equiv
-j_0y\exp\left(\frac{-x^2-y^2}{2a^2}\right),\quad
j_y=j_0x\exp\left(\frac{-x^2-y^2}{2a^2}\right). \label{curr}\ee

It follows from Eq. (\ref{curr}) that there is the probability
current in the region $x, y\sim a$ but its divergence is equal to
zero everywhere. Note that $|\Psi_0({\bf r})|^2$ and
two-dimensional current (\ref{curr}) satisfy the continuity
equation \bb \frac{\partial|\Psi_0({\bf r})|^2}{\partial t}+
\frac{\partial j_x}{\partial x}+\frac{\partial j_y}{\partial y}=0
\label{cont}\ee everywhere. It follows from here that
$|\Psi_0({\bf r})|^2$ is conserved in time. Thus, the state
(\ref{ground}) carries but not creates electrical charge. In a
definite sense one can say that the localized electron becomes
``movable'' in the presence of magnetic field.

Results can be treated quasi-classically. Let two-dimensional
vector field ${\bf j}$ be a complex quantity ${\bf j}=j_x+ij_y$ in
which components $j_x$ and $j_y$ are functions of the complex
variable $z=x+iy$. Note that ${\bf j}$ is a solenoidal vector.
Computing ${\bf c}=[\mathbf{\nabla\times j}]$, we obtain \bb
[\mathbf{\nabla\times j}]=\frac{\partial j_y}{\partial
x}-\frac{\partial j_x}{\partial
y}=\frac{A}{\pi}\left(1-\frac{zz^*}{d^2}\right)\exp\left(-\frac{zz^*}{d^2}\right),
\label{rot} \ee where $A=e\hbar/ma^4,\quad d^2=2a^2$ and $z^*$ is
the complex conjugate of $z$. The field point in which ${\bf c}\ne
0$ is the vortex point or the field vortex. One can write ${\bf
j}$ in the form \bb {\bf j}=\frac{i I(z)}{2\pi z^*},\label{inten}
\ee where \bb
I(z)=Azz^*\exp\left(-\frac{zz^*}{d^2}\right)\label{int1} \ee is
the vortex intensity that is the vector field circulation on a
closed contour encircling the vortex located in the point $x, y=0$
(see, Figure 1).

\begin{figure}[h]
\vspace{-0.2cm} \centering
\includegraphics[scale=0.3]{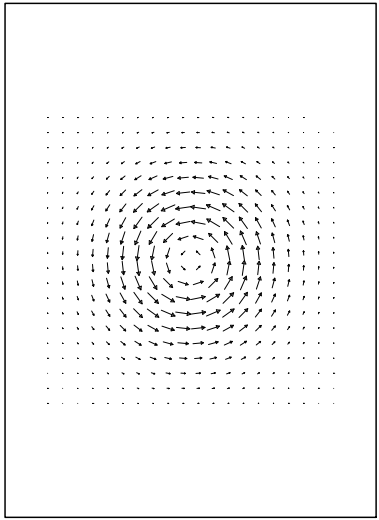}
\caption{Probability current} \vspace{-0.1cm}
\end{figure}

It should be borne in mind that for three-dimensional fields
$I(z)$ defines only the vortex projection on the normal direction
to the plane $z=0$, while the vortex is three-dimensional vector.
For planar fields the vortex can be considered as a vector
perpendicular to the plane $z=0$.

If there are $s$ attractive $\delta$-potentials located in points
$x_k, y_k$ then the current vector is equal to the sum \bb {\bf
j}=\frac{i}{2\pi}\sum\limits_{k=1}^{s}\frac{I(z_k)}{z-z_k}.\label{tot1}
\ee

It is interest to compare function (\ref{ground}) with the
analogous wave function of an electron in a single attractive
$-\lambda\delta({\bf r})$ potential in the absence of magnetic
field. The latter has been easily found in the form \bb\Psi({\bf
r})=\frac{\lambda
C}{2\pi}K_0\left(\sqrt{\frac{2m|E_0|(x^2+y^2)}{\hbar^2}}\right),
\label{ground1} \ee where $C$ is constant, $K_0(z)$ is the
MacDonald function of $0$th order, $E_0$ is the electron energy.
Now electron is localized in the region
$l_0\sim\sqrt{\hbar^2/2m|E_0|}$. The probability current densities
for state (\ref{ground1}) are equal to zero everywhere. Hence, the
state (\ref{ground1}) does not carry electrical charge.

\section{Resume}

We have shown that the effect of magnetic field on localized
electron states leads to appearing the probability current in some
region as well as to enlarging (for a number of physical models) the
localization region of the electron in the plane perpendicular to
the magnetic field.

For the effective mass $m=m_e$ the ratio \bb \frac{a}{l_0}\cong
4\cdot 10^2\sqrt{\frac{|E_0|}{B}}, \label{ratio}\ee where $|E_0|$
and $B$ should be taken in the electron-volt(s) and kilo-gauss(s),
respectively. For typical cases $|E_0|\cong 0.1\div 0.5$ eV,
$B\cong 1\div 50$ kG. The time, which an electron is in
three-dimensional potential well of the width $r_0\ll l_0$ or
$r_0\ll a$ is related to the one out of the well as
$r_0^2\ln^2(r_0/l_0)/l_0^2$ without or $r_0^2/a^2$ in the presence
of magnetic field.

Using the model under discussion one can see a main result of
work \cite{chib} concerning the decreasing  of the number of
defects in thin defect oxide films after the magnetic field
treatment.

\end{document}